\documentclass[aps,twocolumn,english,superscriptaddress,showpacs,bibliography]{revtex4-1}
\usepackage{amsmath}
\usepackage{amssymb,scalerel}
\usepackage{graphicx}
\usepackage[colorlinks,bookmarks=false,citecolor=red,linkcolor=blue,urlcolor=blue]{hyperref}
\usepackage{lipsum, color}
\usepackage{slashed}
\usepackage{bbm}
\usepackage{bbold}
\usepackage{amsmath,mathtools}
\usepackage{amssymb}
\usepackage{wasysym}
\usepackage{soul}
\usepackage{multirow}
\usepackage{xspace}
\usepackage{bm}
\usepackage{dsfont}
\usepackage[htt]{hyphenat}
\usepackage{float} 
\usepackage{placeins}
\usepackage{makeidx}
\usepackage[framemethod=default]{mdframed}
\usepackage{showexpl}
\usepackage{etoolbox}
\usepackage{lineno}

\allowdisplaybreaks
\date{\today}
\begin{document}
\title{Super universality of dimerised $SU(N+M)$ spin chains}
\author{A. M. M. Pruisken}
\email{a.m.m.pruisken@contact.uva.nl}
\affiliation{Institute for Theoretical Physics, Science Park 904, 1098 XH Amsterdam, The Netherlands}
\author{Bimla Danu}
\email{bimla.danu@physik.uni-wuerzburg.de}
\affiliation{Institut f\"ur Theoretische Physik und Astrophysik and W\"urzburg-Dresden Cluster of Excellence ct.qmat, Universit\"at W\"urzburg, 97074 W\"urzburg, Germany}
\author{ R. Shankar}
\email{shankar@imsc.res.in}
\affiliation{The Institute of Mathematical Sciences, C I T Campus, Chennai 600 113, India}
\begin{abstract}
We explore the  physics of the quantum Hall effect using the Haldane mapping of dimerised $SU(N+M)$ spin chains, the large $N$ expansion and the density matrix renormalization group technique. We show that while the transition is first order for $N+M >2$, the system at zero temperature nevertheless displays a continuously diverging length scale $\xi$ (correlation length). The numerical results for $(M, N) = (1,3), ~ (2, 2),~(1, 5)$ and $(1, 7)$ indicate that $\xi$ is a directly observable physical quantity, namely the spatial width of the edge states. We relate the physical observables of the quantum spin chain to those of the quantum Hall system (and, hence, the $\vartheta$ vacuum concept in quantum field theory). Our numerical investigations provide strong evidence for the conjecture of super universality which says the dimerised spin chain quite generally displays all the basic features of the quantum Hall effect, independent of the specific values of $M$ and $N$. For the cases at hand we show that the singularity structure of the quantum Hall plateau transitions involves a universal function with two scale parameters that may in general depend on $M$ and $N$. This includes not only the Hall conductance but also the ground state energy as well as the correlation length $\xi$ 
with varying values of $\vartheta \sim \pi$.
\end{abstract}
\maketitle

\section{Introduction}  In this paper, we are addressing several long standing issues  relating to the robustness of the integer quantum Hall plateaus as well as the concept of super universality in the theory of dimerised quantum spin chains. How is the physics $SU(N)$ spin chains, the main focus in this paper, related to the physics of the quantum Hall effect (QHE)?  The answer to this question is twofold. First, using replica field theory, Levine, Libby and Pruisken  showed that the physics of the QHE can be inferred from the grassmannian $U(N+M)/U(N)\times U(M)$ sigma models with a topological term ($\vartheta$ angle)~\cite{LevineLibbyPruisken1983, LevineLibbyPruisken84, Girvin1990, PruiskenMPB2010, abrahams201050}. Second, Haldane has shown that the low energy physics of the $SU(2)$ spin chain can be mapped onto that of an $SU(2)/U(1)$ sigma model~\cite{Haldane1983_physlett, Haldane1983}. In this special case with $M=N=1$ the angle $\vartheta$ only takes on the value $=0$ or $\pi$ depending on whether the spin $S$ is an even or odd multiple of $\frac{1}{2}$. In later work~\cite{Affleck1985} it was argued that the mapping can be generalized to include $SU(N+M)$ spin chains with a dimerisation parameter  which we term $\epsilon$. This leads to the same grassmannian sigma model with a parameter $\vartheta$ that varies continuously with varying values of $\epsilon$.

The above derivations \cite{Haldane1983_physlett,Haldane1983, Affleck1985} are
valid in the $S\rightarrow\infty$ limit. However, they have technical problems
that complicate a systematic $1/S$ expansion of the parameters of the field
theory. An alternative derivation detailed in Ref.~\cite{Pruisken2008}
resolves these problems for the $SU(1+1)$ case and formulates a well defined
$1/S$ expansion. These results are easily generalized to the $SU(N+M)$ case
and will be presented in a forthcoming publication~\cite{PruiskenBimlaShankar_unpbls}.

It is well known that the sigma model flows toward the strong coupling regime~\cite{BHZ1980, Girvin1990} which is - generally speaking - inaccessible analytically. However, it is also well known that this regime is numerically accessible, namely based on the density matrix renormalisation group (DMRG) approach to quantum spin chains~\cite{White1993, Schollwock2005}. The main purpose of the present paper, therefore, is to use the DMRG as an unequivocal test of the distinctly different strong coupling ideas that over the years have emerged in the study of both $SU(N+M)$ quantum spin chains~\cite{Pruisken2005, Pruisken2008} and the closely related grassmannian $U(N+M)/U(N)\times U(M)$ sigma model~\cite{PruiskenJTP2009,  PruiskenMPB2010}.

Quite surprizingly, the fundamental significance of the ``massless chiral edge excitations" in the problem of QHE or equivalently, the ``dangling edge spins" in the quantum spin chain, has not been fully appreciated so far. These edge excitations are important not only in the definition of the transport coefficients (``conductances"~\cite{Pruisken1999_3, BSkoric1999, Pruisken2005, Pruisken2008}) but also in our conceptual understanding of more general issues in quantum field theory, notably the quantization of topological charge, the existence of robust topological quantum numbers etc.~\cite{Burmistrov07,  PruiskenMPB2010}. %

Advances along these lines have ultimately led to the resolution of old controversies that have spanned the subject to date. We mention, in particular, the ``large $N$" picture of
the $\vartheta$ vacuum concept~\cite{DADDA197863, EWritten1979, SidneyColeman1976, Affleck_7_1980, Affleck_8_1980} as opposed to the distinctly different ``instanton" picture~\cite{{Callan1979, Rajaraman1987, Polyakov1987}}. However, these different views have in fact turned out to be complementary~\cite{PruiskenJTP2009, PruiskenMPB2010}.  This new development is consistent with the original critique by A. Jevicki~\cite{Jevicki} and is diametrically opposite to what historically has been dubbed ``the arena of bloody controversies"~\cite{Coleman1985}. 

It is clear that the conflicting views in quantum field theory have had a dramatic impact on the development of a microscopic theory of the quantum Hall effect. Besides the {\em robust quantization} which emerged as a new and unexpected feature of the $\vartheta$ angle, the idea of a {\em continuously diverging} correlation length $\xi$ - describing the quantum Hall plateau transition - has been a particularly difficult stumbling block for many years~\cite{Affleck19851, Affleck19882, AffleckIan1991}. These otherwise well established experimental phenomena were believed to be ``incompatible"~\cite{Ver-85, Affleck19851, Affleck19882} with the general views based on the ``large $N$" picture which indicated, amongst many other things, that the transition at $\vartheta = \pi$ is a {\em first order} one, for all values $M + N >2$. Such beliefs have, in turn, set the stage for incorrect mathematical claims and ideas in the literature such as the ``failure" of the replica method, the ``superiority" of supersymmetric representations etc.~\cite{Ver-85, Zirnbauer1994, Zirnbauer1997}.

More recently the large $N$ steepest descend methodology of the $CP^{N-1}$ model has been revisited~\cite{PruiskenJTP2009,  PruiskenMPB2010}. Unlike the prevailing expectations in the field it was shown that the physics of the QHE is, in fact, displayed by the $\vartheta$ vacuum concept in general, for all non-negative values of $M$ and $N$. This naturally leads to the idea of {\em super universality} of topological principles which means that mathematical issues such as the replica limit only play a role of secondary significance.

In what follows we shall further explore and extend the super universality concept based on the DMRG 
simulations of the open spin chain with an odd number of spins and finite $M$ and $N$.
The numerical data for the ground state energy with varying system size ($L$) and dimerization ($\epsilon$) are being compared with the large $N$ approach to spin chains~\cite{AffleckPRBlargen1988, ReadSachdev1989} as well as grassmannian sigma models~\cite{PruiskenJTP2009,  PruiskenMPB2010}. The DMRG results now serve as a direct check on the general expectation in the field which says that the transition at 
$\epsilon =0$ (or $\vartheta = \pi$) is a first order one, for $M+N>1$.

The primary focus of this paper, however, is on the physics of the ``edge" the details of which are difficult to obtain analytically.
We are specifically interested in the magnetic properties of spin chains since they can be used as a probe for the ``massless chiral edge excitations" in the problem~\cite{Pruisken1999_1, Pruisken1999_2, Pruisken1999_3, BSkoric1999} or, equivalently, the ``dangling edge spins"~\cite{Pruisken2005, Pruisken2008}.  The DMRG data now indicate that the spatial width or ``penetration depth" of the edge excitations diverges continuously as $\epsilon$ approaches the critical value (or, equivalently, as the angle $\vartheta$ approaches $\pi$). This directly measurable physical quantity is naturally identified with the correlation length $\xi$ of the system. 

 Next, there is the problem of extracting the most fundamental quantity from DMRG, namely the Hall conductance $\sigma_H$ with varying $L$ and $\epsilon$. Generally speaking, this kind of computation  is complicated since it demands an explicit knowledge the ``bulk" excitations and those of the ``edges"~\cite{PruiskenBurmistrov05,  PruiskenMPB2010}. However, by making use of the dual symmetry of the spin chain, along with the macroscopic conservation law for the magnetization, one can introduce an alternative definition of the linear response formula that is suitable for DMRG purposes. This permits a numerical study of the robustness of the QHE as well as the critical singularities of the quantum Hall plateau transition.

In the last part of this paper we propose an extended version of the super universality concept that includes the critical behaviour of not only the Hall conductance but also the ground state energy as well as the correlation length. After a simple re-scaling of the numerical data with $M + N > 2$ we find that the singularity structure can be expressed in terms of a single universal function $F(X)$ where $X$ generally denotes for the linear dimension of the system, i.e. it stands for either $L$ or $\xi$.

\section{Dimer model} 
\begin{figure}[htbp]
\centering
\includegraphics[width=0.4\textwidth]{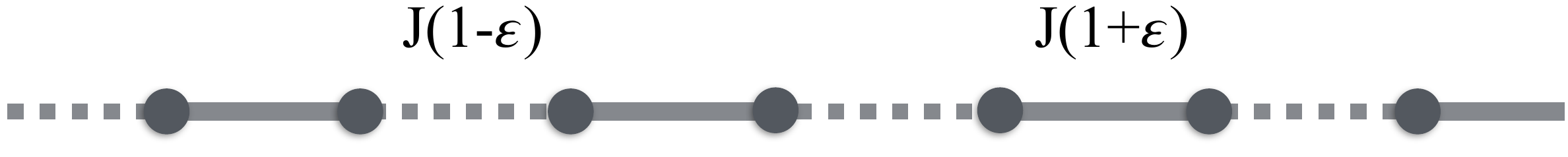}%
\caption {Interacting dimerised spin chain.}
\label{dimerised_spin_chain} 
\end{figure}

Introducing a dimerisation parameter $\epsilon$ defined by assigning couplings $J(1+\epsilon)$ and $J(1-\epsilon)$ to adjacent spin-pairs (see Fig.~\ref{dimerised_spin_chain}), we can write the Hamiltonian of the $SU(M+N)$ spin chain as follows
\begin{equation}
\mathcal{H} = J \sum_{j}  \sum^{M+N}_{\alpha,\beta=1}\left\{ \big(1+(-1)^j \epsilon\big) \hat{\bf S} ^{}_{{2 j},\alpha \beta} \cdot  \hat{\bf S} ^{}_{{2 j + 1},\beta \alpha}\right\} .
\label{Hamil}
\end{equation}
Here, the spin operators satisfy the commutation relation $[ {{\hat {\bf S}}^{}}_{\alpha\beta},  {{\hat {\bf S}}^{}}_{\mu\nu}]= \delta^{}_{\mu\beta}\hat{{\bf S}}^{}_{\alpha\nu}- \delta^{}_{\alpha\nu} {\hat {\bf S}}^{}_{\mu\beta}$. They are in the spin-1/2 representation of $SU(M+N)$ with the little group $U(M)\times U(N)$. 

\noindent{Our} main focus is on the {open} spin chain with {edges}. This displays an obvious {\em dual symmetry} $\epsilon \leftrightarrow -\epsilon$, provided the total number of sites is {odd}, say $2 N_s +1$. The effective action of this system is quite simple in the limit where $M$ is finite and $N \rightarrow \infty$ or vice versa. For example, the ground state involves $N_s$ disconnected {\em dimers} with a total energy 
\begin{equation}\label{E0}
\mathrm{E}_0 (\epsilon) = -N_s e_2 (1+ |\epsilon|) 
\end{equation}
where $e_2$ is the dimer energy for coupling $J$~\cite{AffleckPRBlargen1988, ReadSachdev1989}. This result is precisely the same for a {\em closed}
system with $2 N_s$ sites, indicating that a {\em first order} quantum phase transition takes place when the parameter $\epsilon$ goes through zero. The sole difference, however, is that a ``dangling spin" appears at the ``edge" of an {\em open} chain. The effective action now involves the solid angle $\Omega[V]$ of the $SU(M+N)$ matrix variable $V$
\begin{eqnarray}
\Omega[V] = \int_0^\beta dt ~\mathrm{tr} V \partial_t V^\dag \Lambda~~~;~~~\Lambda=\left(
\begin{matrix}
\mathbb{1}_M & 0 \\
0 & - \mathbb{1}_N
\end{matrix}
\right)
\end{eqnarray}
Assume that for $\epsilon < 0$ the dangling spin is located at the edge on the left ($j =0$); similarly, for $\epsilon > 0$ the single spin appears at the edge on the right ($j = L = 2 N_s$). The complete effective action for the open spin chain in the large $N$ limit now equals
\begin{equation} \label{Sopen-largeN} 
S_{open} \rightarrow \mathcal{F}_0 + S_{edge} [V] ~ .  
\end{equation}
Here, $\mathcal{F}_0 = \beta \times \mathrm{E}_0 (\epsilon)$ denotes the ``bulk" free energy and $S_{edge} [V]$ describes the critical theory of the ``edge" 
\begin{eqnarray}
 S_{edge} [V] &=& i \frac{k(\epsilon)}{2} \Omega[V(L)] +i \frac{1-k(\epsilon)}{2} \Omega[V(0) ] 
\label{Sedge-largeN}
\end{eqnarray}
with $k(\epsilon) =\frac{1}{2}\left(1+\frac{\epsilon}{|\epsilon|}\right)$ the Heaviside step function.

\subsection{ Haldane mapping} Next, to make contact with the theory of the %
non-linear sigma model we write
\begin{eqnarray}
 S_{edge} [V] %
&=& \frac{i}{2} \Omega[V(0)] + 2 \pi i k(\epsilon) \mathcal{C} [\mathcal{Q}]. %
\label{Sedge-largeN-top}
\end{eqnarray}
Here, 
$\mathcal{Q}=V^\dag \Lambda V \in \frac{U(M+N)}{U(M)\times U(N)}$ has a fractional topological charge $-\frac{1}{2} < \mathcal{C} [\mathcal{Q}] \le \frac{1}{2}$. Expressed as a two dimensional space-time integral we have

\begin{eqnarray}
\mathcal{C} [\mathcal{Q}] %
&=& \frac{1}{16 \pi {i} } \int d^2 x~
\mathrm{tr} \in_{\mu\nu} \mathcal{Q} \partial_\mu \mathcal{Q} \partial_\nu \mathcal{Q}. 
\end{eqnarray}

which must be compared with the more general results of linear response theory. Similar to the dimer model, the purpose of the general theory is to formulate an effective action for the ``edge" modes $V$ or $\mathcal{Q}$ relative to a theory of ``bulk" excitations. The latter always corresponds to a compact space-time geometry or, for that matter, a closed spin chain. Specifically, we must compare
\begin{eqnarray}
 2 \pi i k(\epsilon) 
\mathcal{C} [\mathcal{Q}] &\leftrightarrow& \frac{\sigma_0}{8} \int d^2 x~
\mathrm{tr} \left( \partial_\mu \mathcal{Q} \right)^2 +
2 \pi i \sigma_H \mathcal{C} [\mathcal{Q}] .
\label{All-action}
\end{eqnarray}
From replica field theory of the electron gas we know that $\sigma_{H} =\sigma_{H} (\epsilon;\beta,L)$  precisely stand for the Kubo formulae for the macroscopic  ``longitudinal" and ``Hall" conductance respectively.  Apparently, the spin chain with $N\rightarrow \infty$ displays the quantum Hall effect, i.e.
\begin{eqnarray} \label{qHe-dimer-1}
\sigma_0 (\epsilon; \beta, L) \leftrightarrow 0 ~~~;~~~
{\sigma_H}(\epsilon; \beta, L) 
&\leftrightarrow& k(\epsilon)
\end{eqnarray}
for all $\beta$ and $L$. 
At the same time, one can probe the critical edge excitations more directly by measuring the local magnetization 
$\mathcal{M}_j \propto \mathrm{tr} \langle {\hat {\bf S}_j} \Lambda \rangle$. Eq.~\eqref{Sedge-largeN} implies
\begin{equation}\label{edge-magn-largeN}
{ \mathcal{M}_{j=0} =  \frac{1}{2} \left(1 - k(\epsilon) \right) 
~~;~~~
\mathcal{M}_{j=2N_s} =  \frac{1}{2} k(\epsilon) } .
\end{equation}
Notice that sum $\sum_j \mathcal{M}_j = \frac{1}{2}$ is a conserved quantity as it should be. Eq.~\eqref{edge-magn-largeN} now indicates that at criticality, a spin $\frac{1}{2}$ gets transported over macroscopic distances, from one edge of the spin chain to the other.

In conclusion, one can say that the dimer model provides a very simple but profound demonstration of the super universal features of the $\vartheta$ vacuum concept that are inaccessible otherwise.   

\section{Instantons}
Provided $N$ and $L$ are finite, one always finds that the discontinuity in Eqs.~\eqref{E0} and \eqref{qHe-dimer-1} gets smoothed out due the tunneling events (instantons) between the two different dimer states. The situation is completely analogous to the recently revisited large $N$ expansion of the $CP^{N-1}$ model or $SU(N)/U(N-1)$ non-linear sigma model~\cite{PruiskenMPB2010}. In brief, the dimensionless ground state energy $\mathcal{E}_g=2 {\mathrm{E}_0 (\epsilon)}/{L e_2}$ with $L=2N_s$ the length of the spin chain, can in general be written as follows
\begin{eqnarray}
\mathcal{E}_g =  -d -b \sqrt{\epsilon^2+m^2 (L)}.
\label{eg_def-1}
\end{eqnarray}
Here, $b,d = 1 +\mathcal{O}(\frac{1}{M+N})$ whereas the function $m(L)$ is the most significant quantity. It has the general form
\begin{eqnarray}\label{mL_def-1}
m(L) = L^{-\alpha} e^{-\beta L -\gamma}
\end{eqnarray}
with positive coefficients given by
\begin{equation}\label{mL_def-2}
\alpha = 1~~;~~\beta = \frac{(MN)^{3/2}}{2(M+N)^2}~~;~~\gamma = \frac{1}{2} \ln\left( \frac{MN}{4} \right). 
\end{equation}

\subsection{Phase diagram}
It is important to keep in mind that Eqs.~\eqref{eg_def-1} and~\eqref{mL_def-1} apply to {\em closed} spin chains without edges. Nevertheless, in what follows we will argue the instanton results explain most of the important features of {\em open} systems 
as well.
For example, one might expect that the quantity $m(L)$ in Eq.~\eqref{eg_def-1} primarily affects the {\em critical regime} 
$|\epsilon| \lesssim m(L)$ where the discontinuities in $\mathcal{E}_g$ and $\sigma_H$ are smoothed out. On the other hand, the effect of $m(L)$ is negligible 
or exponentially small in the {\em quantum Hall regime} $|\epsilon| \gtrsim m(L)$ where Eq.~\eqref{qHe-dimer-1} is likely to be valid. There are, in fact, two important conclusions that one can draw at this stage. 
\begin{figure}[h]
\centering
\includegraphics[width=0.4\textwidth] {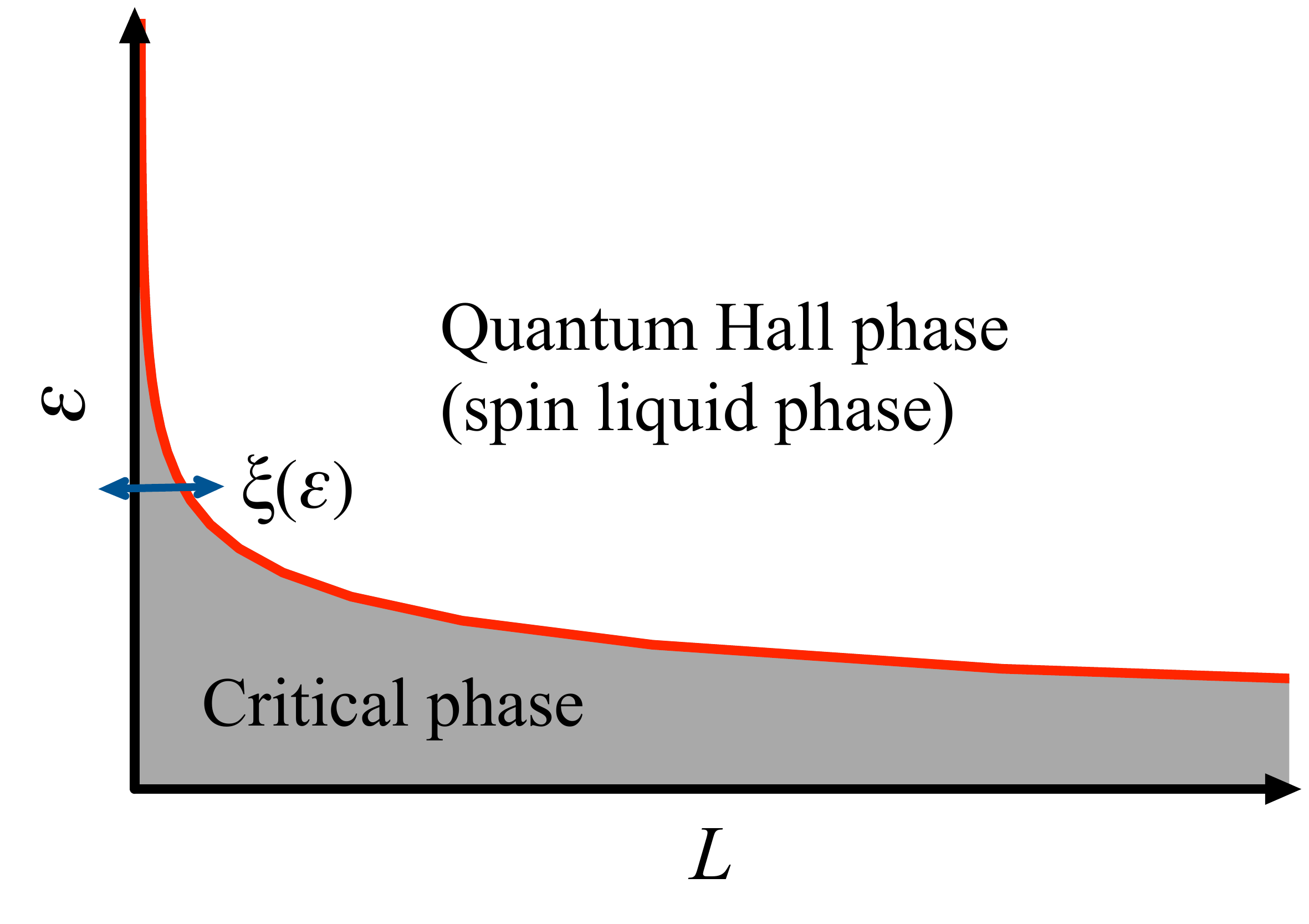}%
\caption {Sketch of 
$|\epsilon| = m (L)$ (solid red line) with $m(L)$ given by Eqs.~\eqref{mL_def-1} and \eqref{mL_def-2}. This line defines the correlation length $\xi (\epsilon)$ which diverges continuously as $\epsilon \rightarrow 0$, see text.}
\label{Epsilon-vs-Edgewidth-1}
\end{figure}
\begin{enumerate}
\item
The line $|\epsilon| = m(L)$ represents the interface between the two distinctly different physical regimes in the problem, see 
Fig.~\ref{Epsilon-vs-Edgewidth-1}. It is natural to identify the function 
\begin{equation}\label{xi}
\xi(\epsilon) = m^{-1} (\epsilon) 
\end{equation}
as the {\em correlation length} in the problem. Remarkably, $\xi (\epsilon)$ diverges in a continuous fashion as $\epsilon$ passes through zero. 

\item Exact expressions for the response parameters $\sigma_0$ and $\sigma_H$ in Eq.~\eqref{All-action} have been obtained {based} on the closely related large $N$ steepest descend methodology of the $CP^{N-1}$ model~\cite{PruiskenMPB2010}. Specifically, one measures the response of the system to a change in boundary conditions (BC), namely in going from periodic BC in space-time to an open system or free BC. In this special case, the response can be very simply expressed in terms of ordinary derivatives of the ground state energy $\mathcal{E}_g$ of the closed system (with periodic BC). Given the anisotropic space-time geometry at hand we now have 
$\sigma_{xx} \ne \sigma_{tt}$ and we can write
\begin{eqnarray}
\sigma_{tt} &=& 0 ;~~\sigma_{xx} = -  m(L) \frac{\partial \mathcal{E}_g}{\partial m(L)}  = \frac{m^2 (L)}{\sqrt{\epsilon^2 + m^2 (L)}} \nonumber \\
\sigma_H &=& \frac{1}{2} \left[ 1- \frac{\partial \mathcal{E}_g}{\partial b\epsilon} \right] = \frac{1}{2} \left[ 1+ \frac{\epsilon}{\sqrt{\epsilon^2 + m^2 (L)}} \right] .
\label{sigma-H}
\end{eqnarray}
\end{enumerate}
\section{Numerical objectives}
Fig.~\ref{Epsilon-vs-Edgewidth-1} and, in particular, Eq.~\eqref{xi} immediately reveal what happens to the ``dangling edge spin" of the open chain as one approaches the critical point. For example, imagine that $L \ge 0$ in Fig.~\ref{Epsilon-vs-Edgewidth-1} actually stands for the position along a semi-infinite spin chain. Then it is natural to assume that for a given value of $\epsilon > 0$ the phrase ``critical phase" really means ``massless chiral edge excitations" that are spread out over the region $0 \le L < \xi(\epsilon)$. So rather than being confined to the single lattice site at the edge ($L=0$) - as naively expected on the basis on the large $N$ result of Eq.~\eqref{Sedge-largeN-top} - the edge excitations are carried in practice by a whole range of spins and eventually, as $\epsilon$ approaches zero, by the entire spin chain. The divergent correlation length $\xi(\epsilon)$ is therefore the {\em mechanism} that enables the dangling spin to travel over macroscopic distances, from one edge of the spin chain to the other, as $\epsilon$ passes through the critical point. In different words, the divergent length scale $\xi(\epsilon)$ {\em unifies} the different types of quantum critical behaviour in the problem, notably the {\em two dimensional} quantum critical behaviour at $\epsilon=0$ (or $\vartheta=\pi$) on the one hand, and the {\em one dimensional} massless edge excitations that generally exist when $\epsilon \ne 0$ (or $\vartheta \ne \pi$) on the other. These previously unrecognized features of the $\vartheta$ angle concept are clearly important, especially given the fact that the quantum phase transition at $\vartheta=\pi$ (or $\epsilon = 0$) is expected to be a {\em first order} one.

Equally remarkable and important, however, is that the diverging correlation length $\xi(\epsilon)$ now manifests itself as a {\em directly observable} physical quantity of the system. This is so because the local magnetization of the groundstate, $\mathcal{M}_j (\epsilon)$, is in fact a probe for the massless excitations that propagate along the edges. However, unlike the large $N$ result of Eq.~\eqref{edge-magn-largeN}, one expects that $\mathcal{M}_j (\epsilon)$ is now spread out over the entire range $0 \le j < \xi(\epsilon)$ rather than the single lattice site at $j=0$ alone. In what follows we embark on the numerical investigations of the aforementioned physical scenarios associated with dimerized spin chains.

\section{DMRG results and discussion} 
Our numerical studies  are based on the DMRG technique~\cite{White1993, Schollwock2005}. Our main focus is on the ground state properties of the open dimerised $SU(N+M)$ spin chain with $S=\frac{1}{2}$ and an odd number of sites. We have used both  infinite  and  finite system DMRG algorithms and constructed the super block configuration with one exact site in the middle of adjacent blocks at each iteration. For each different set of $M$ and $N$ we use a different number ($p$) of most probable eigenstates of the reduced density matrix. Specifically, $p\approx1200$ for $N=2, M=2$, $p\approx 512$ for $N=3, M=1$, $p\approx400$ for $N=5, M=1$ and $p\approx216$ for $N=7, M=1$. To the convergence of the ground state we have performed  8 to 10 sweeps and obtained  our results with the  density matrix  truncation errors of order $\sim 10^{-6} - 10^{-7}$.

\subsection{Ground state energy} 
In Fig.~\ref{Eg_vs_EpsilonL_fixedNM} we plot the %
data for the dimensionless ground state energy $\mathcal{E}_g$ versus $\epsilon$,
\begin{equation}
\mathcal{E}_g = E_0 / N_s e_2
\end{equation} 
where $e_2=J\times MN (1+\frac{1}{M+N})$ denotes the exact dimer energy for coupling strength $J$~\cite{PruiskenBimlaShankar_unpbls}. The results for different values of $(M,N)=(1,3),~(1,5),~(1,7)$ and $(2,2)$ and a range of values of $L=9,~25,~51$ and $111$ compare very well with the large $N$ expression of Eq.~\eqref{E0}. The main difference is the smoothening of the discontinuity at $\epsilon=0$ valid for all finite values of $M, N$ and (odd) $L$. Moreover, for each given value of $L$, the data with increasing $N$ clearly display a tendency toward the dimer model result $\mathcal{E}_g = -1-|\epsilon|$. Therefore, the main features of Fig~\ref{Eg_vs_EpsilonL_fixedNM} are all in remarkable qualitative agreement with the instanton expressions of Eqs.~\eqref{eg_def-1} and \eqref{mL_def-1}.

To account for the fact that the DMRG data and the large $N$ results of Eqs.~\eqref{eg_def-1}-\eqref{sigma-H} really belong to two different physical systems (open versus closed) we mention the following.

\begin{enumerate}
\item
{\em Critical phase.}
The instanton expressions generally provide a good description of the numerical data in the critical phase 
$|\epsilon| \lesssim m(L)$. Specifically, we obtain a good fit using Eqs.~\eqref{eg_def-1} and \eqref{mL_def-1} with the coefficients $b$ and $d$ given by

\begin{equation}\label{b1d2}
b\sim 1-\frac{b_1}{N+M} ~~;~~~ d\sim 1+\frac{d_2}{(N+M)^2}
\end{equation}
and with $b_1$ and $d_2$ positive constants. In addition to this, we find that the function of $m(L)$ in Eq.~\eqref{mL_def-1} applies to the DMRG data as well. Specifically, we employ the general expression $m(L) = L^{-\alpha} e^{-\beta  L+\gamma}$ with $\alpha,~\beta$ and $\gamma$ serving as $M,N$-dependent fitting parameters. It should be mentioned that any detailed comparison of the DMRG data with the numerical values of Eq.~\eqref{mL_def-2} is moot. This is so because the physical mechanisms describing the function $m(L)$ are generally very different dependent on whether one considers an open spin chain rather than closed systems.
\begin{figure}[htbp]
\centering
\includegraphics[width=0.48\textwidth] {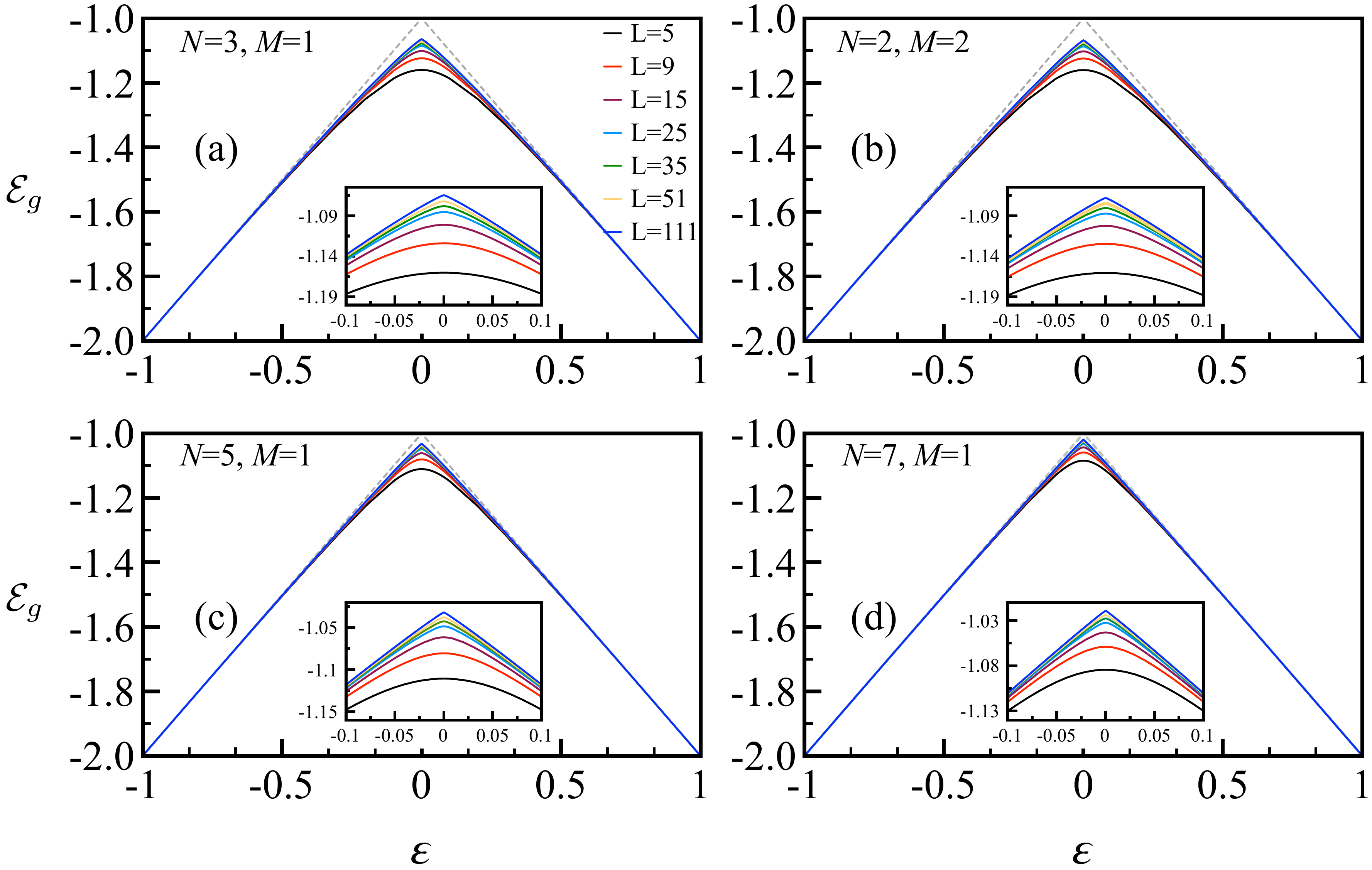}%
\caption {DMRG results for the dimensionless ground state energy ($\mathcal{E}_g$) {\it vs.} dimerisation parameter ($\epsilon$) for different $N,M$ values and system sizes ($L$). The dashed lines represent the large $N$ saddle point result of Eq. \eqref{E0}. }
\label{Eg_vs_EpsilonL_fixedNM}
\end{figure}
\item
{\em Spin liquid phase.} Similar conclusions apply to the $CP^{N-1}$ results for the conductances. For example, if one inserts the DMRG data for $\mathcal{E}_g$ in Eq.~\eqref{sigma-H} then one would naively conclude that the  spin chain does not display the quantum Hall effect. The problem, of course, is that the subtle features of the edges are being mishandled since the $\mathcal{E}_g$ in Eq.~\eqref{sigma-H} is really defined for closed systems. 
\end{enumerate}
We now proceed and embark on the magnetic features of the open spin chain. Along the way we will find an alternative definition of the Hall conductance and obtain a better insight into the physical properties of both the critical and spin liquid phases of the problem.

\subsection{Massless chiral edge excitations}
\begin{figure}[htbp]
\centering
\includegraphics[width=0.49\textwidth]{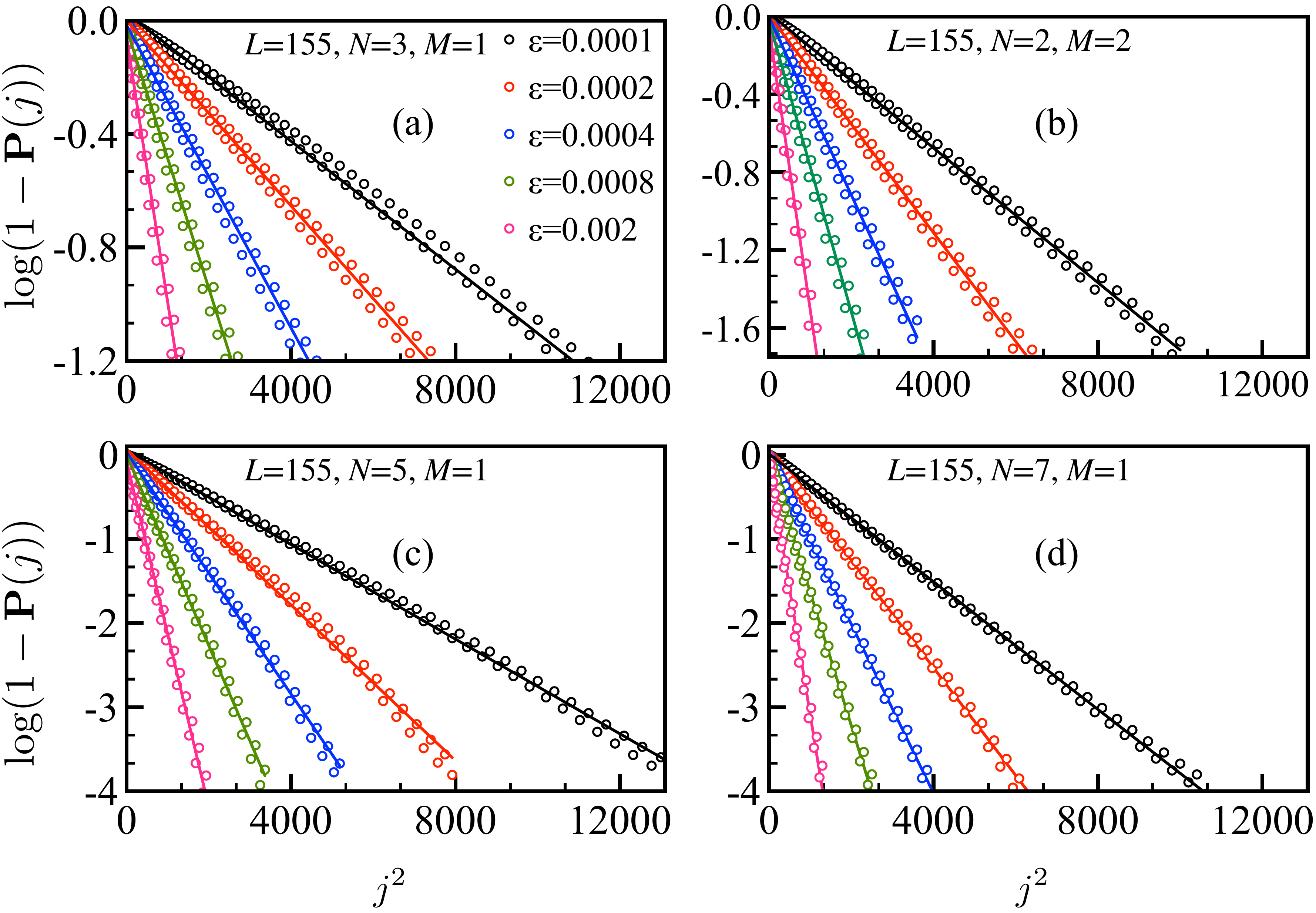}
\caption {DMRG data of $\log(1-\mathrm{\bf P}(j))$ {\it vs.} $j^2$, see text.  The solid lines represent the optimal fitting based on Eq.~(\ref{Prop}).} 
\label{Mi_vs_i_SUN_L155}
\end{figure}

To define the penetration of the edge excitations into the interior of the system it is convenient to introduce the ``cumulative" edge magnetization
\begin{equation}\label{Pj}
\mathrm{\bf P} (j)= \sum_{i=0}^j \mathcal{M} (i) 
\end{equation}
with $j \in \{0, 1, 2, \dots , 2N_s \}$ denoting the lattice site.  Since the total magnetization is a conserved quantity we use the normalization $\mathrm{\bf P} (L) = \sum_{i=0}^L\mathcal{M} (i) =1$ with $L=2N_s$. 
In Fig.~\ref{Mi_vs_i_SUN_L155} we plot the numerical data for $(1-\mathrm{\bf P} (j))$ versus $j^2$ on a log-linear scale for different values of $\epsilon$ close to the critical point. Discarding the anti-ferromagnetic fluctuations in $\mathrm{\bf P} (j)$ - which are relatively small - we have
\begin{equation}\label{Prop}
\mathrm{\bf P} (j) = 1 -  A \times \exp{\left\{ - B {j}^2 \right\}}
\end{equation}
with $A,B>0$, provided $L$ is large enough. It is clear that Eq.~\eqref{Prop}
really stands for the probability of finding the dangling edge spin somewhere in the region of the lattice sites between $j$ and $0$. Fig.~\ref{Mi_vs_i_SUN_L155} clearly indicates that $B=B(\epsilon)$ generally decreases as $\epsilon$ goes to zero. It is therefore natural to identify the quantity $\xi(\epsilon) =1/\sqrt{B}$ as the continuously diverging {\em correlation length} of the system. 

\subsection{Hall conductance}
\begin{figure}[htbp]
\centering
\includegraphics[width=0.48\textwidth]{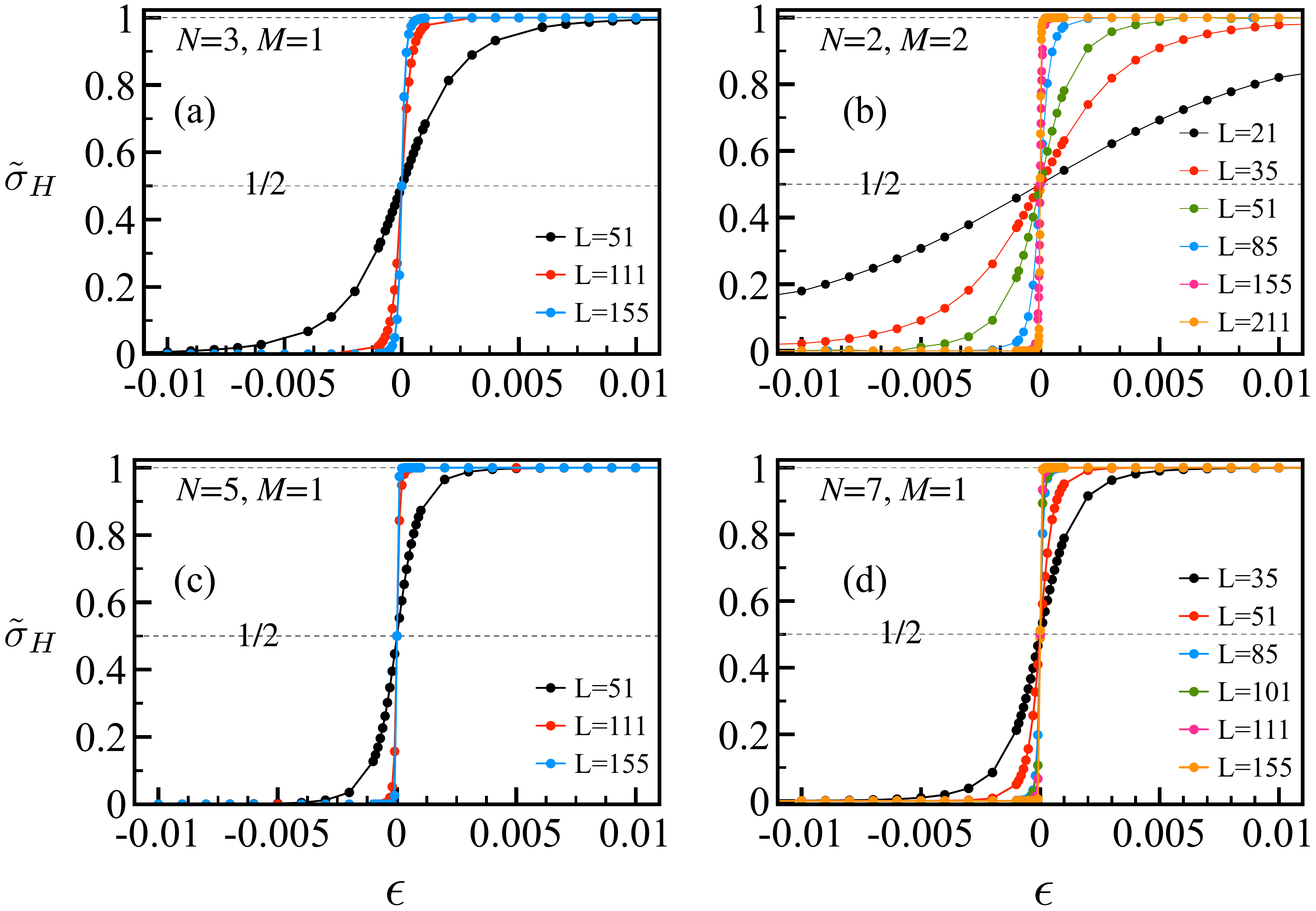}
\caption {DMRG results for the Hall conductance 
($ \tilde{\sigma}_H $) {\it vs} dimerisation parameter ($\epsilon$) for different $N,M$  values and system sizes ($L$).}
\label{SUN-Hall_vs_epsilon0}
\end{figure}

\begin{figure}[htbp]
\centering
\includegraphics[width=0.45\textwidth] {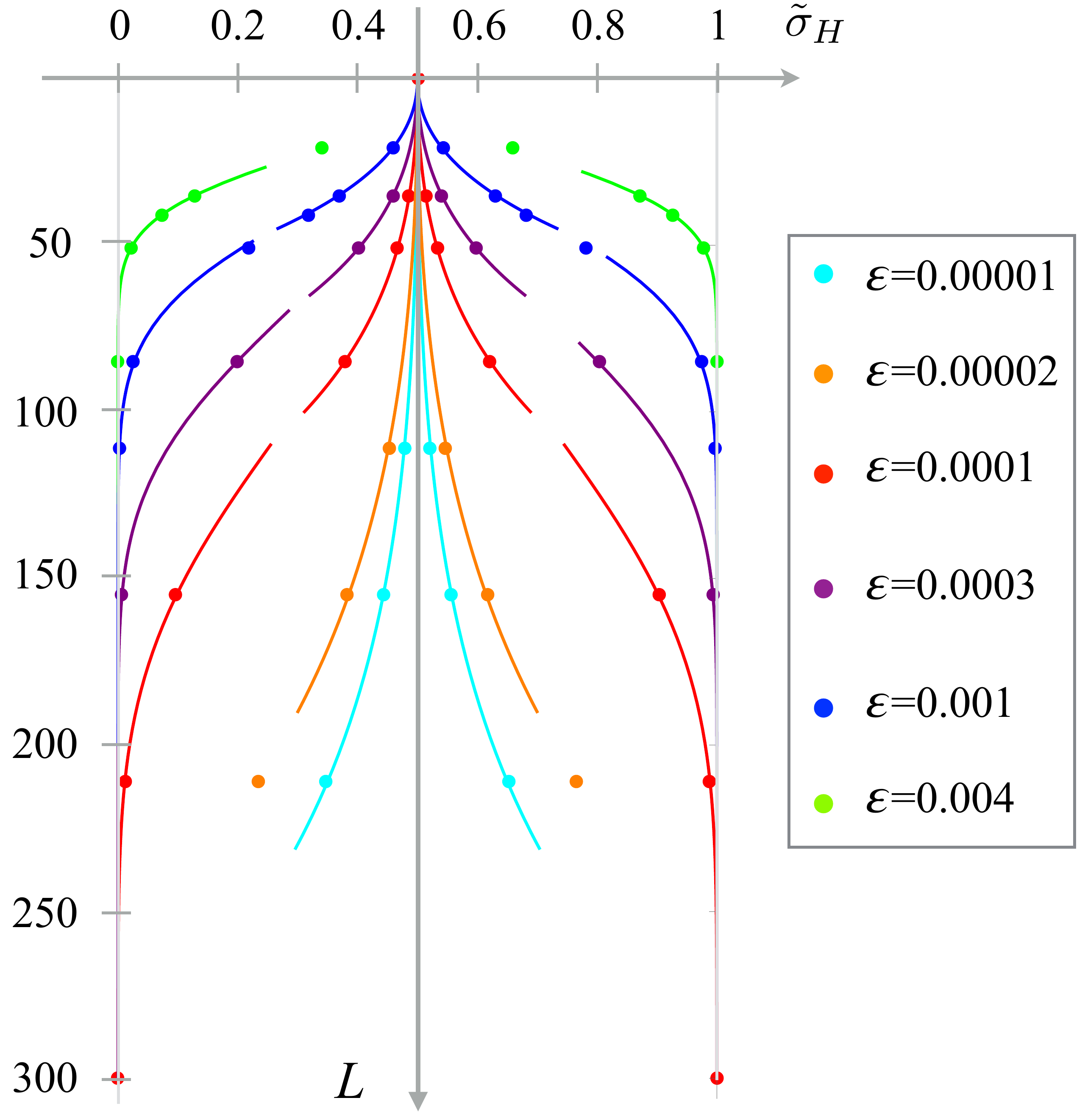}%
\caption {DMRG results for the Hall conductance 
($\tilde{\sigma}_H$) {\it vs} system size ($L$) for $N=M=2$ and different values of the dimerisation parameter ($\epsilon$). The solid lines with $0.3 \lesssim \tilde{\sigma}_H \lesssim 0.7$ are the best fit based on the function $\tilde{\sigma}_H = \frac{1}{2} \pm \Gamma_1 \exp\{\beta_1 L + 2.1 \ln L\}$ with positive coefficients 
$\Gamma_1 \propto |\epsilon|$ and $\beta_1$. 
Those with $ \tilde{\sigma}_H \lesssim 0.3$ and $\tilde{\sigma}_H \gtrsim 0.7$ have been obtained using 
$\tilde{\sigma}_H = \Gamma_2 \exp\{-\beta_2 L^2\}$ and 
$\tilde{\sigma}_H = 1- \Gamma_2 \exp\{-\beta_2 L^2\}$ respectively, with positive $\Gamma_2$ and $\beta_2$.}
\label{SUN-Hall_vs_L00}
\end{figure}

A remarkable feature of open spin chains in the spin liquid phase 
$|\epsilon| > m(L)$ is that the quantization of the Hall conductance is directly related to the conservation of total magnetization.  This is obviously the case in the large $N$ limit where $m(L)=0$, see Eqs.~\eqref{qHe-dimer-1} and~\eqref{edge-magn-largeN}, but the same is true in general. For example, from Eq.~\eqref{Prop} we infer that $\sigma_H (\epsilon)=1-\mathrm{\bf P} (j)$ provided 
$\xi(\epsilon) < j < L$. It is not difficult to generalize this statement to include the critical phase $|\epsilon| < m(L)$ or $L > \xi(\epsilon)$.
Specifically, if we denote the magnetic definition of the Hall conductance by 
$\tilde{\sigma}_H (\epsilon)$ then we can write
\begin{eqnarray}
\tilde{ \sigma}_H  (\epsilon)= 
1- \frac{1}{2} \left[ \mathrm{\bf P} \left( {N_s} \right) + \mathrm{\bf P} \left( {N_s -1} \right) \right] 
\label{sigma-H-II}
\end{eqnarray}
which is nothing but the total magnetization of exactly the right-half of the spin chain.
Notice that duality of the spin chain 
$\epsilon \leftrightarrow -\epsilon$ implies that  
\begin{equation}\label{dual}
\tilde{\sigma}_H (\epsilon) = 1-~\tilde{\sigma}_H (-\epsilon). 
\end{equation}
This fundamental symmetry, sometimes termed ``particle-hole" symmetry, was originally predicted as a corollary of the renormalization theory of the quantum Hall effect~\cite{Girvin1990}. 
It is clearly recognizable in the numerical data plotted in Figs.~\ref{SUN-Hall_vs_epsilon0}-\ref{SUN-Hall_vs_L00}. 
These DMRG results are very similar to the plots obtained using the large $N$ expression for $\sigma_H$, Eq.~\eqref{sigma-H}. In particular,
Fig. ~\ref{SUN-Hall_vs_epsilon0} indicates that $\tilde{\sigma}_H$ with varying $\epsilon$ approaches the dimer result $k(\epsilon)$ of Eq.~\eqref{edge-magn-largeN} as $N$ increases, keeping $L$ fixed and finite. Therefore, just like the ground state energy $\mathcal{E}_g$ we again find remarkable qualitative agreement with the instanton expression for $m(L)$ in Eq.~\eqref{sigma-H}.

Finally, it is interesting to notice that the numerical data of Fig.~\ref{SUN-Hall_vs_epsilon0} and the flow-lines of Fig.~\ref{SUN-Hall_vs_L00} are akin to the results of the first experiments on quantum criticality in the quantum Hall regime~\cite{WeiPRB1986}. This illustrates the fact that superuniversality has a much broader range of validity than $SU(M+N)$ spin chains, the free electron gas ($M=N=0$) and large $N$ expansions alone. In fact, extensive research over many years has shown that the same basic phenomena are being displayed by entirely different physical systems. The most obvious examples include the electron gas in the presence of the Coulomb interaction~\cite{Burmistrov07, PruiskenMPB2010}
and also the problem of ``macroscopic charge quantization" in the single electron transistor~\cite{BurmistrovPRL2008, BurmistrovPRB2010}. 

\subsection{Singularity structure open chains} 
Of principal interest are three distinctly different functions $F(X)$ that diverge continuously as $X$ goes to infinity. 

\noindent{$\bullet$} The first and most significant of these is the second derivative of the ground state energy $\mathcal{E}_g$ at the critical point, 
\begin{eqnarray} \label{F-E}  
F_{Energy} (L) =  \frac{\partial^2 \mathcal{E}_g} {\partial \epsilon^2} \bigg|_{\epsilon \rightarrow 0} ~ 
\leftrightarrow ~ 
\frac{1}{m (L)}.  
\end{eqnarray}
\begin{figure}[htbp]
\centering
\includegraphics[width=0.4\textwidth] {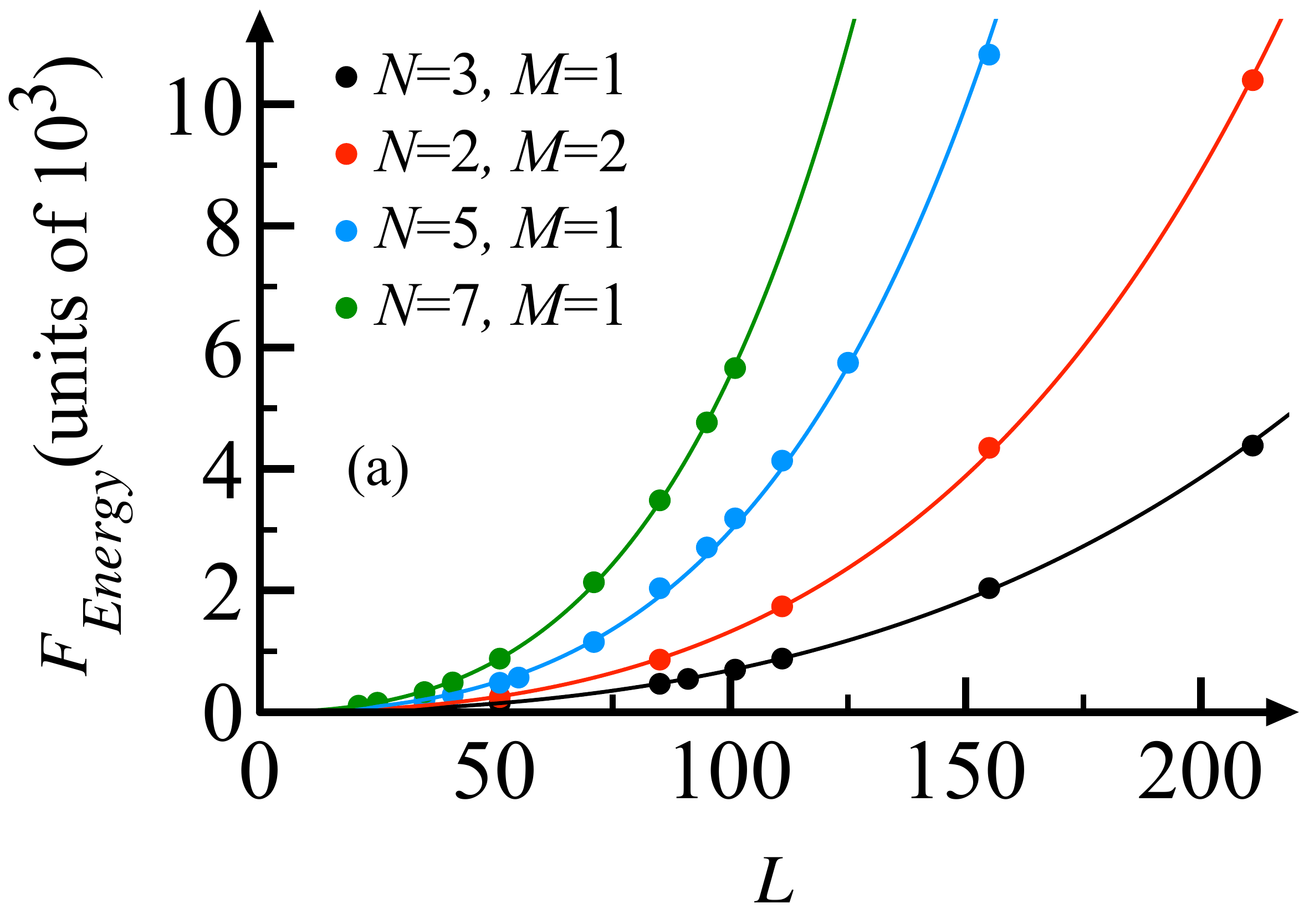}
\includegraphics[width=0.4\textwidth] {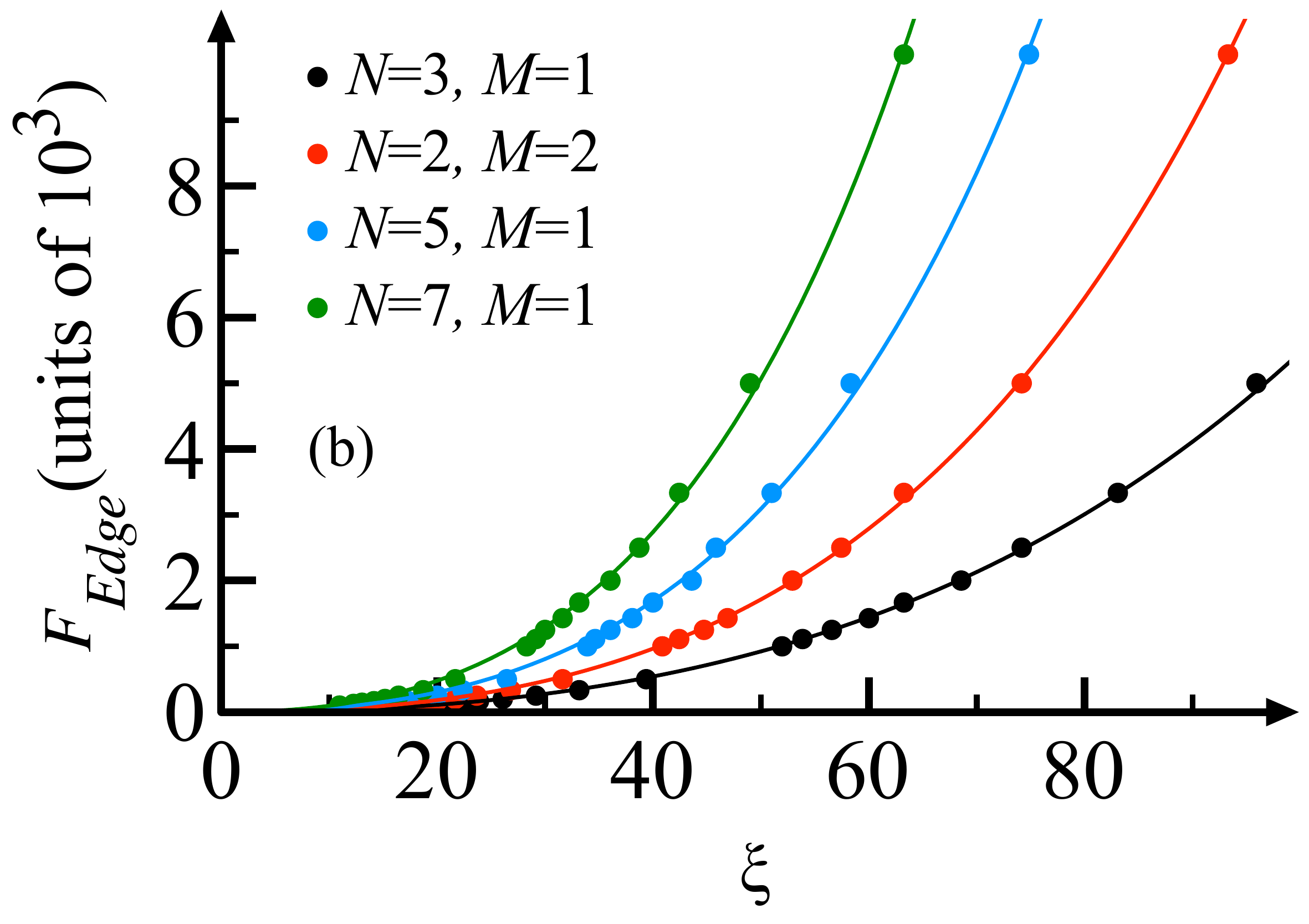}
\caption { DMRG results for the functions $F_{Energy} (L)$ and $F_{Edge} (\xi)$ for different $N,M$ values, see text. The solid lines are optimal fittings based on Eqs.~\eqref{rescaling-i} and \eqref{NEW-F-X}.}
\label{SU_M+N_eng_mag}
\end{figure}

The quantity $1/m(L)$ on the right hand side indicates the instanton result of Eqs.~\eqref{mL_def-1} and~\eqref{mL_def-2} for the closed spin system with varying size $L$. We therefore expect that the DMRG data for $F_{Energy} (L)$ diverge exponentially as $L$ increases.
An explicit demonstration of this divergence can be taken as the experimental proof of a first order quantum phase
transition at $\epsilon = 0$ (or $\theta_B = \pi$).
\newline{The} most practical way of extracting Eq.~\eqref{F-E} from DMRG is to first compute 
$\partial \mathcal{E}_g /\partial \epsilon$ 
for a discrete set of $\epsilon$ values and subsequently determine the second derivative using the standard numerical programs. For the purpose of the present paper, it suffices to fix the spacing $\Delta \epsilon$ 
at $10^{-5}$.
In Fig.~\ref{SU_M+N_eng_mag}(a) we plot the DMRG data sets for four different values of $M$ and $N$.
\begin{figure}[htbp]
\centering
\includegraphics[width=0.4\textwidth] {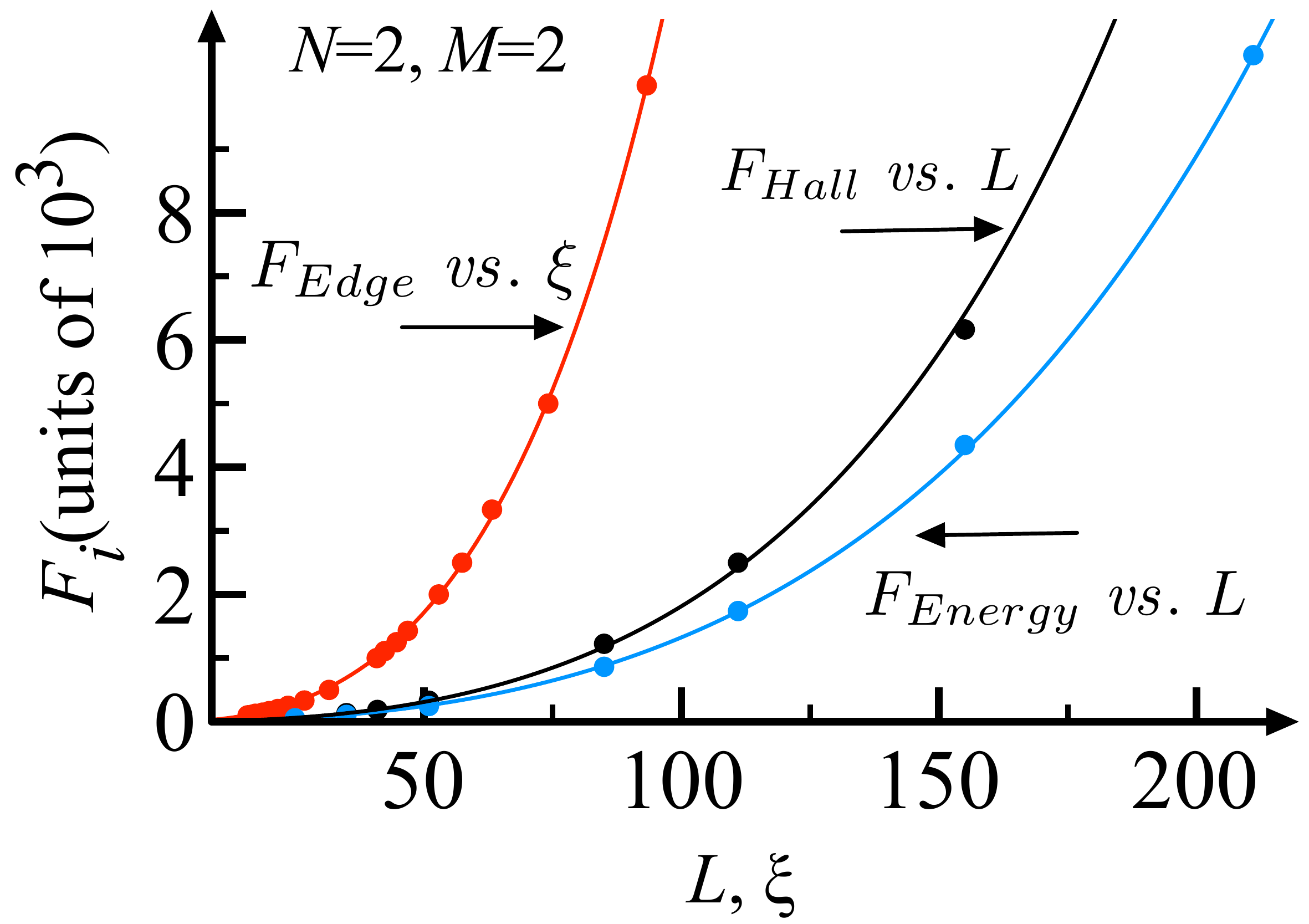}
\caption {DMRG results for the functions $F_{Hall} (L)$, $F_{Energy} (L)$ and $F_{Edge} (\xi)$ with $N=M=2$, see text. The solid lines are optimal fittings based on Eqs.~\eqref{rescaling-i} and \eqref{NEW-F-X}.} 
\label{SU_2+2_eng_mag_hall}
\end{figure}
\noindent{$\bullet$} The second most significant quantity is the correlation length $\xi$ with varying $\epsilon$. We have seen that unlike Eq.~\eqref{F-E}, the $\xi$ is solely defined as a quantity of the ``edge."  As a practical rule we demand that $\xi$ for any given value of $\epsilon$ is determined by the equation
$\mathrm{\bf P} (j=\xi) = 1-e^{-1.5} \approx 0.78\%$ 
where $\mathrm{\bf P} (j)$ is defined by Eq.~\eqref{Pj}.
In words, there is a $78\%$ probability of finding a dangling spin somewhere in the region of lattice sites 
$0 \le j \le \xi$.

In Fig.~\ref{SU_M+N_eng_mag}(b) we plot $1/\epsilon$ versus the DMRG data for $\xi$ for the aforementioned values of $M$ and $N$. The different data sets provide and estimate for the function $F_{Edge} (\xi)$, i.e.
\begin{equation} \label{F-Edge}  
F_{Edge} (\xi ) = \frac{1}{\epsilon}
~~ \leftrightarrow \frac{1}{m(\xi )}  
\end{equation}
where the right hand side is the result of the large $N$ theory, see Eq.~\eqref{xi}. Indeed, the plots of Figs.~\ref{SU_M+N_eng_mag}(a)-(b)
clearly indicate that the functions $F_{Energy} (L)$ and 
$F_{Edge} (\xi)$ display the same qualitative features for varying values of $M$ and $N$.

\noindent{$\bullet$ From the physics point of view, our main interest is obviously in the singularity structure of the quantum Hall plateau transition described by the function

\begin{eqnarray}
\label{FHall_II_vs_epsilon}
F_{Hall} (L) &=& \frac{\partial \tilde{\sigma}_H} {\partial \epsilon} \bigg|_{\epsilon \rightarrow 0}
~~ \leftrightarrow \frac{1}{m(L)}  
\end{eqnarray}
with $\tilde{\sigma}_H$ defined by Eq.~\eqref{sigma-H-II}.
Just like Eq.~\eqref{F-E}, the most practical way of extracting Eq.~\eqref{FHall_II_vs_epsilon} from DMRG is to compute $\tilde{\sigma}_H$ for a discrete set of $\epsilon$ values and subsequently determine the derivative using the standard numerical programs.

In a subsequent paper we will show that the large $N$ limit of Eq.~\eqref{FHall_II_vs_epsilon}, like Eq.~\eqref{F-E}, is $1/m(L)$~\cite{PruiskenBimlaShankar_unpbls}.
For comparison we plot, in Fig.~\ref{SU_2+2_eng_mag_hall}, the DMRG data sets for the three different functions $F_{Energy} (L)$, 
$F_{Edge} (\xi)$ and $F_{Hall} (L)$, taking the case $M=N=2$ as a representative example. Once more, apart from a simple re-scaling of the $X$ and/or $Y$ axis, the results look qualitatively very similar.

\subsection{Universality revealed} 
To find the best solid lines through the data in both Figs.~\ref{SU_M+N_eng_mag} and~\ref{SU_2+2_eng_mag_hall} we are inspired by instanton results of Eqs.~\eqref{mL_def-1} and \eqref{mL_def-2}. Specifically, we write
\begin{equation}\label{rescaling-i}
F_i (X) = a_i F(b_i X) 
\end{equation}
where the subscript $i$ stands for $Energy$, $Edge$ and $Hall$ respectively. The coefficients $a_i$ and $b_i$ are taken as independent fitting parameters and $F(X)$ is an empirical function without free parameters. To fix the thought consider the large $N$ limit with $F_i (X) = 1/m(X)$. We can write
$F(X) = X \exp \left\{{X} \right\}$ in Eq.~\eqref{rescaling-i} and 
\begin{equation}\label{ai-bi}
a_i \rightarrow \frac{N}{M}~~;~~b_i \rightarrow \frac{1}{2} M 
\times \sqrt{\frac{N}{M}}
\end{equation}
keeping in mind that $N \rightarrow \infty$ and $M$ is fixed and finite. Notice that Eq.~\eqref{ai-bi} implies that $F_i (X)$ 
increases as the value of $N$ increases. This feature is in accordance with the DMRG data plotted in Figs.~\ref{SU_M+N_eng_mag} (a) and (b).

The problem, however, is to find an optimal function 
$F(X)$ such that Eq.~\eqref{rescaling-i} can be used for data fitting for all values of $M$ and $N$ as well as the subscript $i$. After a lot of trial and error we found the best function $F(X)$ in Eq.~\eqref{rescaling-i} to be $F(X) = X^{2.1} \exp \left\{{X} \right\}$. We will work with the completely equivalent but more practical expression
\begin{equation}\label{NEW-F-X}
{F}(X) = X^{2.1} \exp \left\{ -2.5+\frac{X}{150} \right\}
\end{equation}
which corresponds, roughly speaking, to the ``average" of the nine different DMRG data sets. 
\begin{figure}[h]
\centering
\includegraphics[width=0.4\textwidth] {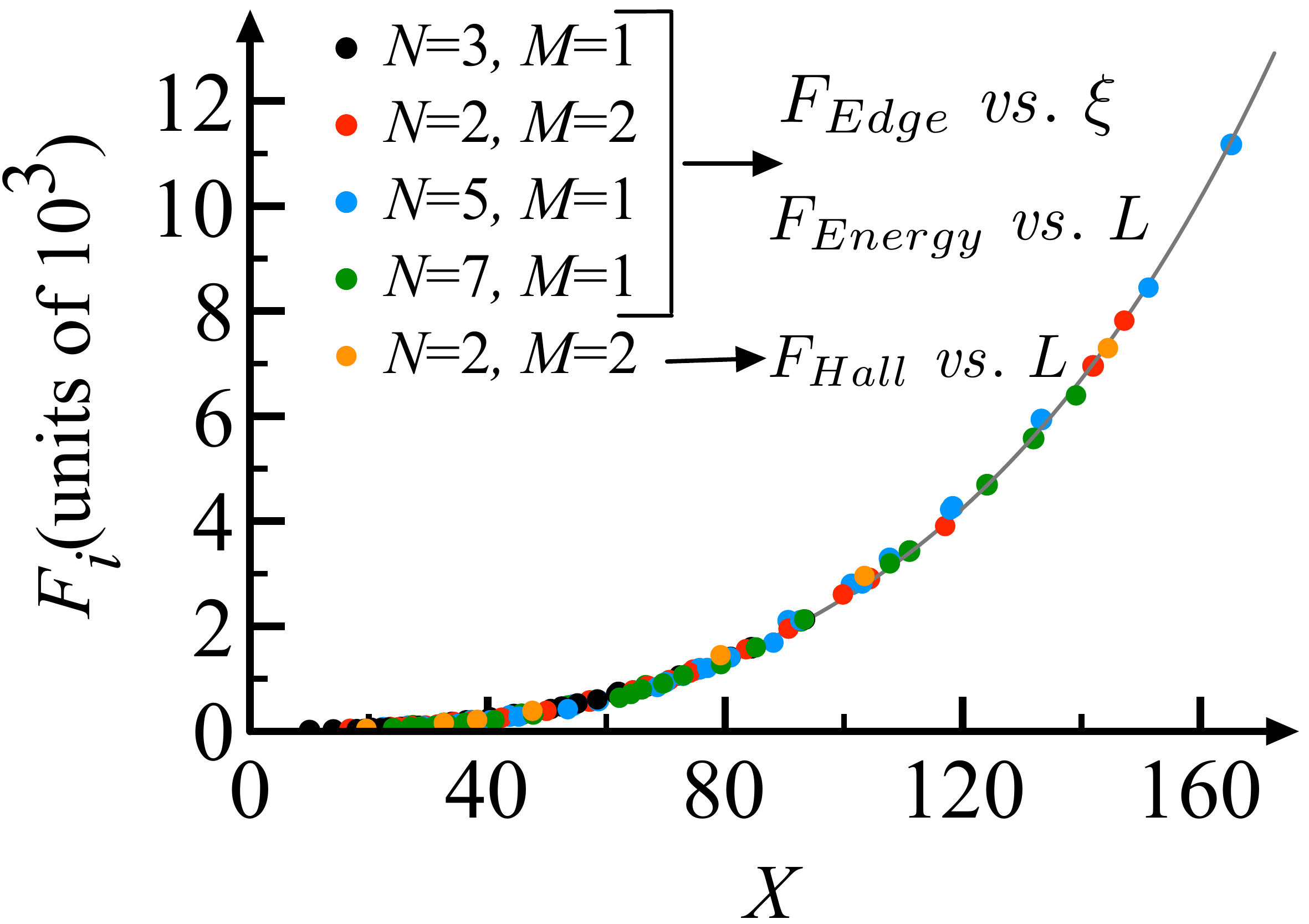}%
\caption {Collapse of the nine different data sets of Figs.~\ref{SU_M+N_eng_mag} and \ref{SU_2+2_eng_mag_hall} after re-scaling. The solid line represents the function ${F}(X)$, Eq. \eqref{NEW-F-X}, with $X= L$ or $\xi$. }
\label{SU_N+M_9data_collapse}
\end{figure}

The solid lines in both Figs.~\ref{SU_M+N_eng_mag} and \ref{SU_2+2_eng_mag_hall} clearly show that  Eqs.~\eqref{rescaling-i} and \eqref{NEW-F-X} fit the DMRG data remarkably well. The most important feature of these expressions, however, is that each of the nine DMRG data sets can be mapped onto the single curve $Y=F(X)$ after a simple rescaling of the $X$ and $Y$ axes. This data collapse is plotted in Fig.~\ref{SU_N+M_9data_collapse}. We see that the re-scaled DMRG data all fall nicely onto the solid line  $F(X)$, as expected.

\section{ Conclusion } 
The principal advancement of this paper is captured in
Fig.~\ref{SU_N+M_9data_collapse}. The DMRG data clearly indicate that the basic predictions of the large $N$ theory can be trusted all the way down to the regime where $M+N$ is of order unity. The results furthermore unify the general  features of a first order phase transition with quantum phenomena that were previously  unrecognized. We mention, in particular, the {\em robustly} quantized Hall plateaus along with the {\em deconfinement} mechanism that facilitates the transport of a dangling spin (or massless edge excitations) over macroscopic distances~\cite{Pruisken2005,Pruisken2008}. This mechanism can be depicted as a spatial separation between the critical  phase and the spin liquid phase (see Fig.~\ref{Epsilon-vs-Edgewidth-1}).

We have refined and extended the concept of super universality to include the correlation length $\xi$ that diverges continuously as one approaches the critical point.
As a result, we can now say that the basic aspects of quantum Hall physics are all generic topological features of the $\vartheta$ vacuum concept on the strong coupling side, for all non-negative values of $M$ and $N$~\cite{LevineLibbyPruisken1983,LevineLibbyPruisken84, Girvin1990, PruiskenMPB2010}. 

It should be mentioned that the space-time geometry of the spin chain is somewhat unnatural, at least as far as the QHE is concerned. The quasi one dimensional geometry with $\beta \rightarrow \infty$ and $L$ finite is clearly very different from the truly two dimensional quantum Hall system which usually involves $\beta \sim L$. Without going into further detail, however, we can say that as an integral aspect of the super universality concept one expects that the basic phenomena are independent of the geometry that one considers. For example, based on the large $N$ theory of the $CP^{N-1}$ model~\cite{PruiskenMPB2010} we know that the differences are solely in the detailed behaviour of the function $F(X)$. Unlike Eq.~\eqref{NEW-F-X} this function is algebraic for systems where $\beta$ and $L$ play a role of equal significance. Specifically,
\begin{equation} \label{NEW-F-X-1}
F(X) \propto X^{1/\nu}
\end{equation}
where $X$ stands for either $L \sim \beta$ or $\xi$. The $\nu$ denotes the correlation length exponent which for the systems at hand is equal to $1/2$, a well known result for a first order transition in two dimensions.

Eq.~\eqref{NEW-F-X-1} is an extremely familiar statement that is valid for both first and second order quantum phase transitions. 
However, as one of the interesting new features of the $\vartheta$ vacuum - and along with that, the quantum Hall plateau transition - we now find that the correlation length $\xi$ {\em always diverges} in a continuous manner as $\vartheta$ approaches $\pi$. Moreover, the physical quantity $\xi$ is {\em directly measurable} and manifests itself  most clearly as a kind of penetration depth of the massless chiral edge excitations. 

Finally, the precise  correspondence between the ``bulk" excitations and those at the ``edges" of the spin chain is actually established the function $F(X)$ as discovered in this paper. This remarkable topological aspect is not seen in the theory of ordinary critical phenomena. 
\acknowledgments   
The authors gratefully acknowledge the HPC Nandadevi cluster at The Institute of Mathematical Sciences for providing computational time.  A.M.M.P. and B.D are indebted to the Institute of Mathematical Sciences (Chennai) for visiting appointments at various  stages of this work.
\bibliography{ref_SUN}
\end{document}